\documentclass[fleqn,usenatbib,letters]{mnras}

\usepackage{newtxtext,newtxmath}

\usepackage[T1]{fontenc}
\usepackage{ae,aecompl}

\usepackage[varg]{txfonts}
\usepackage{natbib,graphicx,amssymb}
\usepackage[english]{babel}
\usepackage{xspace}
\usepackage{lscape}
\usepackage{breqn}
\usepackage{url}
\usepackage{float}
\usepackage{amsmath} 
\usepackage{color}

\newcommand{\sbs}{SBS\,0335$-$052\xspace}
\newcommand{\mum}{$\mu$m\xspace}
\newcommand{\hers}{\textit{Herschel}\xspace}
\newcommand{\spit}{\textit{Spitzer}\xspace}
\newcommand{\hi}{H\,{\sc i}\xspace}
\newcommand{\htwo}{H$_2$\xspace}
\newcommand{\aco}{$\alpha_{\rm CO}$\xspace}
\newcommand{\ha}{H$\alpha$\xspace}
\newcommand{\kms}{km\,s$^{-1}$\xspace}
\newcommand{\msun}{M$_{\odot}$\xspace}
\newcommand{\kkmspc}{K\,km\,s$^{-1}$\,pc$^2$\xspace}
\newcommand{\acounit}{M$_{\odot}$\,pc$^{-2}$\,(K\,km\,s$^{-1}$)$^{-1}$\xspace}

\title[ALMA view of \sbs]{
New ALMA constraints on the star-forming ISM at low metallicity:
A 50 pc view of the blue compact dwarf galaxy \sbs
}

\author[D. Cormier et al.]{
  D. Cormier$^{1}$\thanks{E-mail: dcormier@zah.uni-heidelberg.de},
  G.~J. Bendo$^{2}$,
  S. Hony$^{1}$,
  V. Lebouteiller$^{3}$,
  S.~C. Madden$^{3}$,
  F. Galliano$^{3}$,
  \newauthor
  ~S.~C.~O. Glover$^{1}$,
  R.~S. Klessen$^{1}$,
  N.~P. Abel$^{4}$,
  F. Bigiel$^{1}$, and
  P.~C. Clark$^{5}$\vspace{3pt}
\\
$^{1}$
Institut f\"ur theoretische Astrophysik, 
Zentrum f\"ur Astronomie der Universit\"at Heidelberg, 
Albert-Ueberle Str. 2, 69120 Heidelberg, Germany\\
$^{2}$
Jodrell Bank Centre for Astrophysics,
School of Physics and Astronomy,
The University of Manchester,
Oxford Road, Manchester M13 9PL, UK\\
$^{3}$
Laboratoire AIM, CEA/DSM - CNRS - Universit\'e Paris
  Diderot, Irfu/Service d'Astrophysique, CEA Saclay, 91191
  Gif-sur-Yvette, France \\
$^{4}$
University of Cincinnati, Clermont College, Batavia, OH, 45103, USA\\
$^{5}$
School of Physics and Astronomy, Queen's Buildings, The Parade,
Cardiff University, Cardiff, CF24 3AA, UK
}

\date{Accepted 2017 February 24. Received 2017 February 23; in original form 2016 December 2}

\pubyear{2017}

\begin{document}
\label{firstpage}
\pagerange{\pageref{firstpage}--\pageref{lastpage}}
\maketitle

\begin{abstract}
Properties of the cold interstellar medium of low-metallicity
galaxies are not well-known due to the faintness and extremely
small scale on which emission is expected.
We present deep ALMA band\,6 (230\,GHz) observations
of the nearby, low-metallicity ($12+\log(\rm{O/H})=7.25$)
blue compact dwarf galaxy \sbs at
an unprecedented resolution of 0.2\,arcsec (52\,pc).
The $^{12}$CO\,$J$=2$\rightarrow$1 line is not detected
and we report a 3-$\sigma$ upper limit of
$L_{\rm CO(2-1)}=3.6\times10^4$\,\kkmspc.
Assuming that molecular gas is converted into stars
with a given depletion time, ranging from 0.02 to 2\,Gyr,
we find lower limits on the CO-to-H2 conversion factor
\aco in the range $10^2 - 10^4$\,\acounit.
The continuum emission is detected and resolved over
the two main super star clusters.
Re-analysis of the IR-radio spectral energy distribution
suggests that the mm-fluxes are not only free-free
emission but are most likely also associated with a cold
dust component coincident with the position of the brightest
cluster. With standard dust properties, we estimate its mass
to be as large as $10^5$\,\msun.
Both line and continuum results suggest the presence of
a large cold gas reservoir unseen in CO even with ALMA.
\end{abstract}




\section{Introduction}
\label{introduction}
Star-formation conditions in low-metallicity environments
are not well known due to the lack of observational constraints
on the cold dust and gas.
Recent efforts with ground-based telescopes have
tried to detect the cold gas reservoirs through carbon monoxide
(CO) in nearby low-metallicity galaxies \citep[e.g.,][]{schruba-2012,
elmegreen-2013,cormier-2014,hunt-2015,rubio-2016}.
CO emission is found to be weak in those galaxies
while active, bursty star formation is often evident.
This would imply significantly different star-formation efficiencies
unless CO no longer traces the bulk of the star-forming gas.
Star formation taking place in the atomic gas
\citep[e.g.,][]{glover-2012,krumholz-2012b} as well as
changes in the standard CO-to-\htwo conversion factor
between high- and low-metallicity galaxies have been
suggested \citep[see the review by][]{bolatto-2013}
though the amplitude of those changes is unknown. 
A complementary probe of the ISM in galaxies comes from
the continuum emission. Recent, sensitive observations
in the far-IR/millimeter (mm) with \hers \citep{pilbratt-2010}
have allowed us to constrain the spectral energy distributions
(SEDs) of many nearby low-metallicity dwarf galaxies and
to derive more robust dust masses \citep[e.g.,][]{remy-2015}.
However, most galaxies with metallicities $\le$1/5\,$Z_{\sun}$
remain undetected in the submm/mm regime, implying that
characterizations of a possible cold ($T_{\rm dust}$<20\,K),
massive dust reservoir are still uncertain.
In an effort to characterize the very low-metallicity regime,
we present new ALMA observations of the dwarf galaxy \sbs.

\begin{figure*}
\centering
\includegraphics[clip,trim=-12mm -2mm 2mm 0,scale=.58]{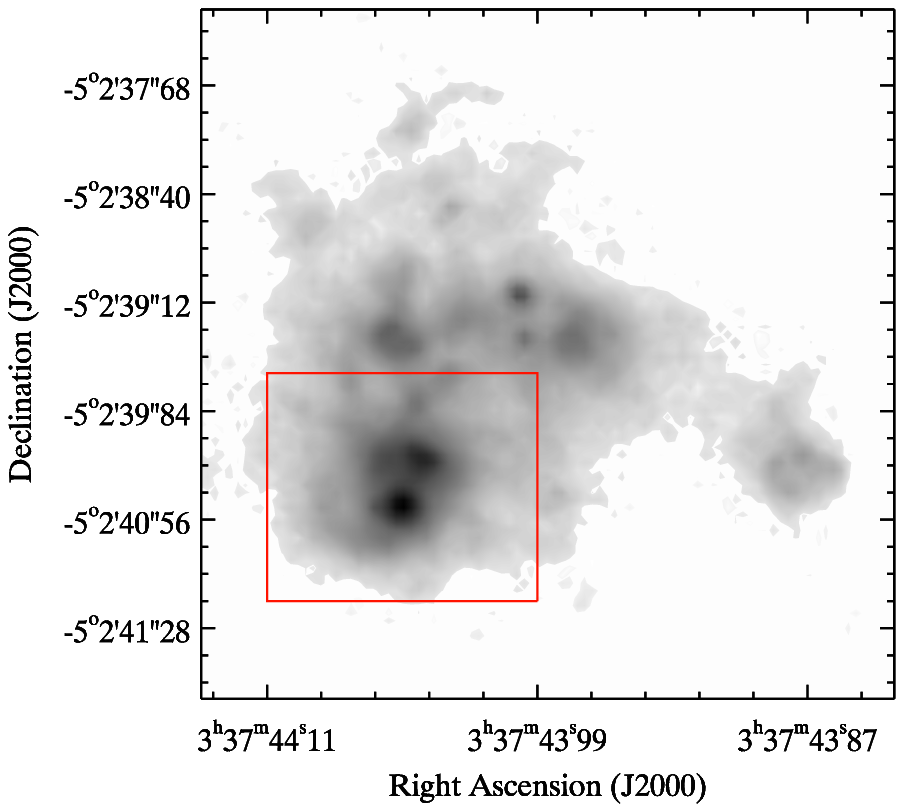}\hfill{}
\includegraphics[clip,trim=-12mm -2mm 2mm 0,scale=.56]{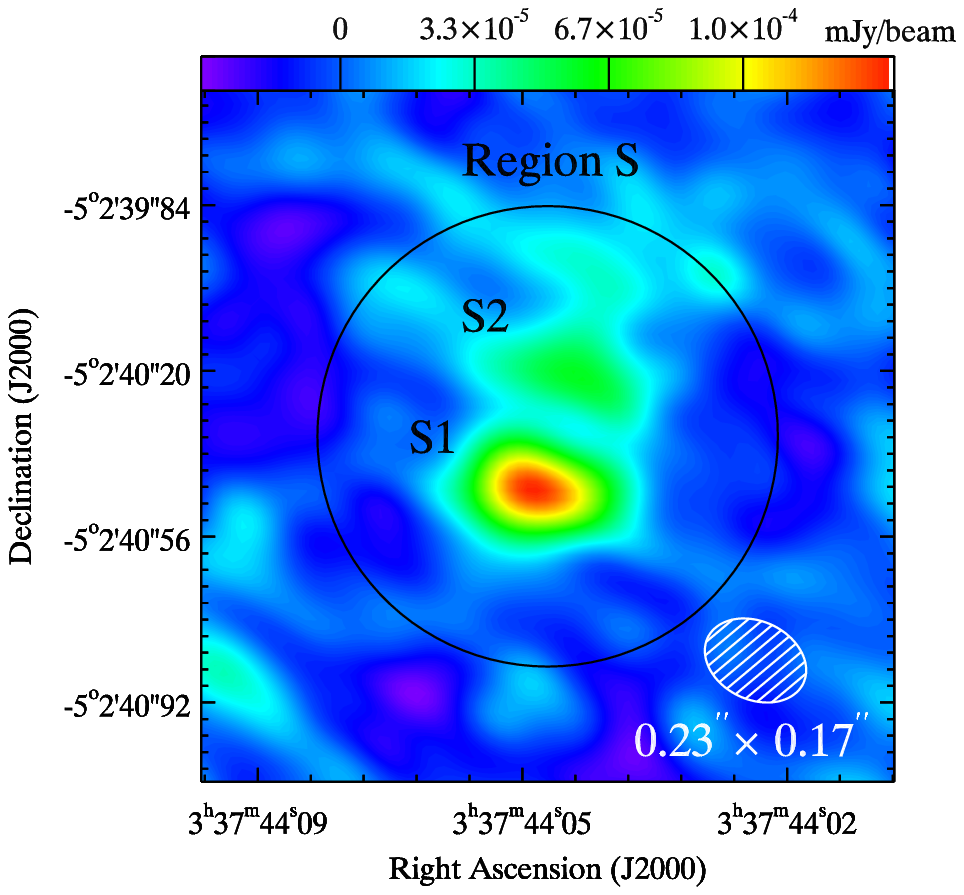}\hfill{}
\includegraphics[clip,trim=3mm 3mm 3mm 5mm,scale=.45]{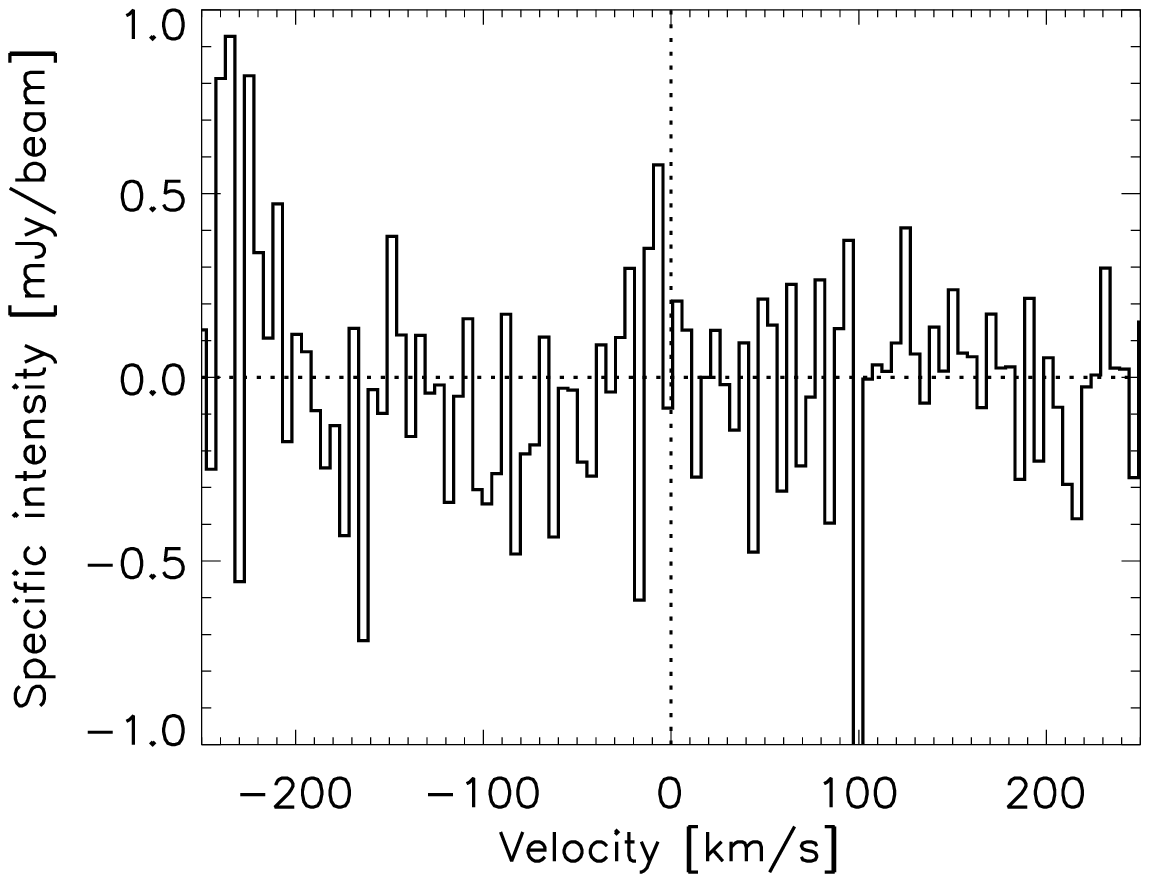}
\vspace{-3pt}
\caption{
\textit{Left:} \textit{HST} ACS FR656N map of the main body of \sbs,
downloaded from the Hubble Legacy Archive (\protect\url{http://hla.stsci.edu/}),
in logarithmic scale. 
The red box indicates the area shown in the middle panel.
\textit{Middle:} ALMA band\,6 continuum image in mJy/beam.
The beam is shown in white. An aperture of radius
0.5\,arcsec encompassing S1 and S2 used for
photometry is shown in black (labeled region S,
as in \protect\citealt{johnson-2009}).
\textit{Right:} ALMA band\,6 CO\,$J$=2$\rightarrow$1
spectrum averaged within one beam centered at the position
of the continuum emission peak (S1).
The spectral resolution is 5\,\kms.
}
\label{fig:fig1}
\end{figure*}

\subsection{General properties of \sbs}
Among nearby low-metallicity galaxies, the blue compact 
dwarf \sbs at 54\,Mpc is one of the best laboratories to study
vigorous star formation in quasi-pristine conditions. 
It is composed of a main body (Fig.~\ref{fig:fig1}) that
we refer to as \sbs.
Its general properties are listed in Table~\ref{table1}.
\sbs is well known as one of the lowest metallicity
galaxies of the nearby Universe, with $12+\log(\rm{O/H})\sim7.25$
\citep{izotov-2009}, and exhibits exceptionally active
star formation, with a total star-formation rate (SFR)
of 0.7\,M$_{\sun}$\,yr$^{-1}$ \citep{remy-2015} and
$\Sigma_{\rm SFR}$ as high as
20\,M$_{\sun}$\,yr$^{-1}$\,kpc$^{-2}$ \citep{johnson-2009}.
Star formation in \sbs is located in 6 compact ($<$60\,pc) 
super star clusters (SSCs; \citealt{thuan-1997}). 
The bulk of its stellar population has an age of less than 500\,Myr. 
The two brightest SSCs, located to the south (S1 and S2 on
Fig.~\ref{fig:fig1}), are the youngest, with ages $\le$3\,Myr
and stellar masses of $\sim10^6$\,\msun \citep{reines-2008} each.

ISO and \spit observations reveal strong mid-IR continuum
from warm dust, no polycyclic aromatic hydrocarbons, and
hard radiation fields \citep[e.g.,][]{thuan-1999,wu-2006}. 
While optical studies measure an average visual extinction $A_{V}$
of  $\sim$0.7\,mag \citep[e.g.,][]{izotov-1997,vanzi-2000,reines-2008},
with values similar in S1 and S2,
dust modeling of the unresolved IR emission yields 
$A_{V}$$\ge$10\,mag \citep[e.g.,][]{thuan-1999,hunt-2005}. 
This discrepancy suggests the presence of a separate component 
with high IR optical depth and probably hidden star formation. 
Although this galaxy remains undetected in CO \citep{dale-2001b,
hunt-2014}, warm molecular gas is present, since near-IR
\htwo lines are detected in the southern clusters \citep[e.g.,][]{vanzi-2000}.

In this letter, we present new ALMA band 6 observations
of \sbs at a resolution of 0.2\,arcsec (52\,pc).
Section~\ref{sect:obs} describes the observations. The
molecular gas reservoir and spectral energy distribution are
analysed in Sections~\ref{sect:ks} and \ref{sect:sed}.
We discuss and summarize our results in Sections~\ref{sect:disc}
and \ref{sect:conc}.

\begin{table}
\caption{Properties of \sbs (E).}
\label{table1}
 \centering
\vspace{-5pt}
  \begin{tabular}{llc}
\hline \hline
    \vspace{-10pt}\\
\multicolumn{3}{c}{General characteristics}\\
\hline
    \vspace{-10pt}\\
Coordinates (J2000)				& 03h37m44.0s, -05d02m40s 			& \\
Distance 						& 54.1\,Mpc						& (*) \\
$12+\log(\rm{O/H})$				& 7.12-7.32						& (1) \\
M$_{\rm stellar}$				& $5.6\times10^6$\,M$_{\odot}$ 		& (2) \\
M$_{\rm HI}$ 					& $4.3\times10^8$\,M$_{\odot}$ 		& (3) \\
SFR 							& 0.7\,M$_{\odot}$\,yr$^{-1}$ 			& (4) \\
$\Sigma_{\rm SFR}$	(region S)		& 20\,M$_{\odot}$\,yr$^{-1}$\,kpc$^{-2}$  	& (5) \\
\hline
    \vspace{-10pt}\\
\multicolumn{3}{c}{ALMA band\,6 observations at 230\,GHz}\\
\hline
    \vspace{-10pt}\\
Continuum (S1) & 0.207$\pm$0.013 (mJy) & \\
Continuum (S2) & 0.088$\pm$0.011 (mJy) & \\
Continuum (region S) & 0.317$\pm$0.048 (mJy) & \\
CO\,$J$=2$\rightarrow$1 & $<$0.41 (mJy/beam)$^{a}$ & \\
\hline \hline
    \vspace{-10pt}\\
\end{tabular}
\begin{minipage}{8.2cm}\small
$^{a}$~RMS of the spectrum at a spectral resolution of 5\,\kms.
(*) Luminosity distance from NED, based on \ha velocity
from \cite{moiseev-2010};
(1)\,\cite{izotov-2009}; (2)\,\cite{reines-2008};
(3)\,\cite{ekta-2009}; (4)\,\cite{remy-2015};
(5)\,\cite{johnson-2009}.
\end{minipage}
\end{table}

\section{Data}
\label{sect:obs}
\subsection{ALMA band\,6 data}
Observations of \sbs were carried out in 5 blocks
between August, 30 2015 and September, 23 2015
as part of the Cycle\,2 program 2013.1.00916.S (PI Cormier).
We targeted the continuum and $^{12}$CO\,$J$=2$\rightarrow$1
line in band\,6 at a spatial resolution of 0.2\,arcsec.
Observations were made in extended configuration
with 35 antennas. The field-of-view is 15\,arcsec.
One spectral window was centered at a rest
frequency of 230.5\,GHz with resolution 1.3\,\kms and
the 3 other spectral windows were centered at
232.5, 245.5 and 247.5\,GHz.
Each window has bandwidth of 1.875\,GHz.
All windows were combined to make the continuum map
(total bandwidth of 7.5\,GHz centered around 239\,GHz).

We reprocessed and recalibrated the raw data
from the archive using the Common Astronomy
Software Applications version 4.6.0 \citep{mcmullin-2007}.
Data from each execution block were first processed
through a set of standard calibration steps to correct
phase and amplitude variations versus time and versus
channel and to flux calibrate the data.
Data from antennas or channels which have irregular
phases or amplitudes are flagged during the calibration.
The final images were then created using the {\sc clean}
algorithm with natural weighting.
The source was calibrated using quasars where the
estimated flux densities have uncertainties of $<\sim$15\,per cent.
The reconstructed Gaussian beam size in the final images
is 0.23\,arcsec$\times$0.17\,arcsec with a position angle of
65$^{\circ}$, which matches the full-width at half maximum
of the axes of the beam before deconvolution as well as
the axis orientation.
The continuum is detected at a signal-to-noise of $10$
towards the peak of emission and resolved into two
compact knots (S1 and S2, see Fig.~\ref{fig:fig1}).
We achieve an rms of 0.011\,mJy/beam in the continuum.
We extract fluxes for the two SSCs by fitting 2D
ellipses constrained by the shape of the ALMA beam.
We fit each SSC at a time, masking out the other SSC.
Emission from S1 is marginally resolved (fitted with
an ellipse of size $1.15\times$ the beam).
Over the region S, we measure a total flux of 0.32\,mJy
(Table~\ref{table1}) with a circular aperture of radius
0.5\,arcsec covering S1 and S2. The sum of the
Gaussians agrees with the total within errors, indicating
no significant extended emission. Uncertainties are
given as the rms per beam times $\sqrt{N_{\rm beams}}$.
The CO\,$J$=2$\rightarrow$1 line is not detected.
We report an rms of 0.41\,mJy/beam, for a spectral
resolution of 5\,\kms.
The maximum recoverable scale of our observations
is $\sim$5\,arcsec ($\sim$1.3\,kpc), so we are not missing
extended emission in our analysis.

\subsection{Ancillary data and star-formation rates}
We compare our ALMA observations to ISM data of \sbs
as well as radio data. We use \hers photometry at a resolution
$>$5\,arcsec from \cite{remy-2015}.
We complement those with ALMA band\,7 (346\,GHz) data
from \cite{hunt-2014} at $\sim$0.6\,arcsec resolution.
Additionally, we use radio continuum data from the VLA
(1.3 to 6\,cm) at $\sim$0.5\,arcsec resolution
and SSC-extracted fluxes from \cite{johnson-2009}.
\sbs was observed at 1.5--22\,arcsec resolution with the VLA
(1.3 to 21\,cm) by \cite{hunt-2004}.
Although \sbs is not resolved in the \hers data or in the VLA data
from \cite{hunt-2004}, there is evidence that S1 and S2 dominate
the global emission at other wavelengths. Hence we consider
emission from region S as total emission.
The total SFR of the galaxy, measured by combining the
\ha and total infrared emission, is $\sim$0.7\,M$_{\sun}$\,yr$^{-1}$
\citep{remy-2015}.
From resolved studies, 1.3\,cm emission yield SFRs of 0.38 and
0.27\,M$_{\sun}$\,yr$^{-1}$ for S1 and S2 \citep{johnson-2009}.

\section{Star formation and molecular reservoir}
\label{sect:ks}

\begin{figure}
\centering
\includegraphics[clip,width=8.2cm]{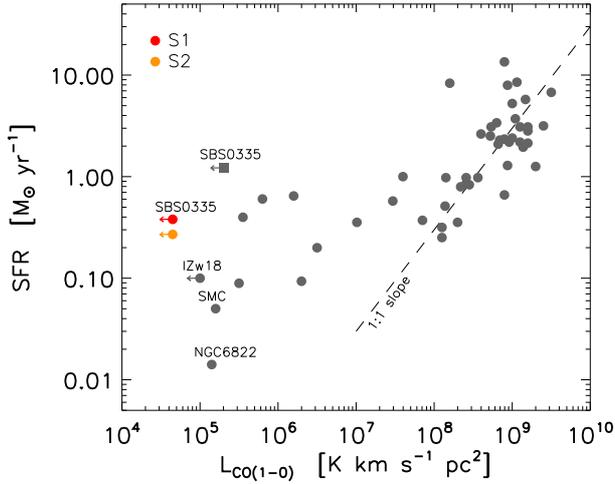}
\vspace{-8pt}
\caption{
Star-formation rate (SFR) as a function of CO\,$J$=1$\rightarrow$0 luminosity.
Limits are 3-$\sigma$.
Literature data are in grey, compiled from:
\protect\cite{schruba-2012} (circles) and
\protect\cite{hunt-2014} (square).
To guide the eye, the dashed line indicates a power-law
slope of 1. \sbs is $10^4$ times fainter in CO
than a solar-metallicity galaxy exhibiting a similar SFR.
}
\label{fig:fig2}
\end{figure}
%
\begin{figure}
\centering
\includegraphics[clip,trim=5mm 12mm -3mm 0,width=8.2cm]{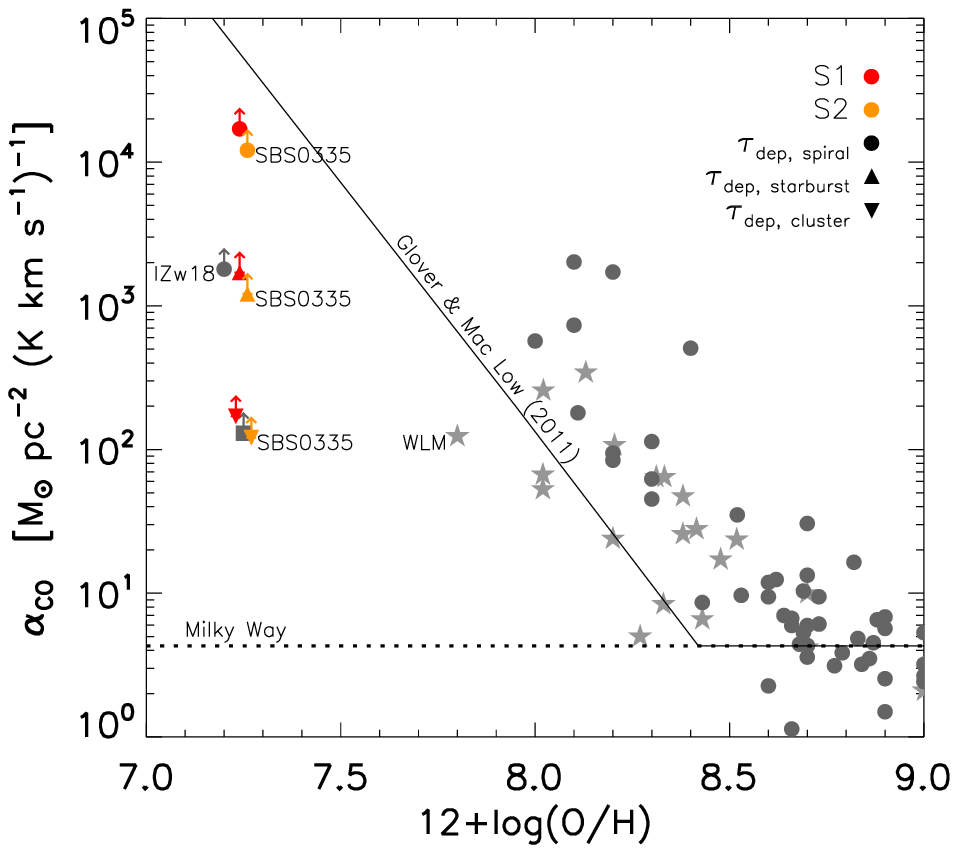}
\vspace{-8pt}
\caption{
Conversion factor $\alpha_{\rm CO}$ as a function of metallicity.
\sbs: circles, upward triangles, and downward triangles
are for $\tau_{\rm dep}$ of 2\,Gyr, 0.2\,Gyr, and 20\,Myr respectively.
Limits are 3-$\sigma$.
Literature data points are in grey, compiled from:
\protect\cite{schruba-2012} which assumes $\tau_{\rm dep}$ of 1.8\,Gyr (circles);
\protect\cite{hunt-2014} for \sbs which assumes a subtended area of
3.5\,arcsec$^2$ and $\tau_{\rm dep}$ of $\sim$7$\times10^7$\,yr (square);
\protect\cite{israel-1997}, \protect\cite{leroy-2011}, and \protect\cite{elmegreen-2013}
for Local Group dwarf galaxies which use dust continuum
emission (stars).
We overplot predictions from \protect\cite{glover-2011} where $A_{V}$
is converted to oxygen abundance assuming a linear scaling
($A_{V}$ of 3.5\,mag corresponds to $12+\log({\rm O/H})=8.42$).
}
\label{fig:fig3}
\end{figure}

The new ALMA observations put stringent constraints
on the amount of CO-emitting molecular gas in \sbs.
Assuming a broad line width of 50\,\kms, as measured
from \hi \citep{ekta-2009}, we derive a conservative
3-$\sigma$ $L_{\rm CO(2-1)}$ limit of $3.6\times10^4$\,\kkmspc.
Our solid angle is $0.044$\,arcsec$^2$.
Taking a CO(2-1)/CO(1-0)
ratio of 0.8 \citep[e.g.,][]{sage-1992}, this corresponds
to a $L_{\rm CO(1-0)}$ limit of $4.5\times10^4$\,\kkmspc.
Previous CO\,$J$=1$\rightarrow$0 and CO\,$J$=3$\rightarrow$2
measurements by \cite{dale-2001b} and \cite{hunt-2014},
with solid angles of 23\,arcsec$^2$ and 1\,arcsec$^2$ and
a CO(3-2)/CO(1-0) ratio
of 0.6, put $L_{\rm CO(1-0)}$ limits of $8.9\times10^6$
and $1.8\times10^5$\,\kkmspc, respectively.
Figure~\ref{fig:fig2} shows the CO\,$J$=1$\rightarrow$0 luminosity
as a function of SFR for \sbs and several nearby galaxies which
measurements are compiled in \cite{schruba-2012}.
Our ALMA data provide a new limit on the CO luminosity in \sbs.
\sbs clearly stands out in being so CO faint for a star-forming galaxy.

The CO-to-\htwo conversion factor (\aco) being unconstrained
at low metallicity, we do not convert this luminosity into a limit on
the mass of \htwo. Instead, we hypothesize that star formation proceeds
with a given depletion time-scale, $\tau_{\rm dep}$=M(\htwo)/SFR.
We assume three different molecular gas depletion times:
2\,Gyr - a common value for normal, disc-type galaxies \citep[e.g.,][]{bigiel-2008};
0.2\,Gyr - a common value for starburst galaxies \citep[e.g.,][]{daddi-2010}; and
20\,Myr - a common value for local molecular clouds \citep[e.g.,][]{lada-2012}.
Significantly reduced depletion time-scales - similar to starburst
values -  have been found by \cite{kepley-2016} for another
dwarf galaxy, II\,Zw\,40.

From the CO luminosity and SSC-derived SFR values,
we can constrain \aco. The \aco limits that we find for
S1 and S2 are shown in Figure~\ref{fig:fig3}
and are around $10^2$\,\acounit for cloud depletion
times (20\,Myr), and $10^3 - 10^4$\,\acounit, amongst the highest,
for whole-galaxy depletion times ($0.2 - 2$\,Gyr).
The most surprising finding is that, even at our 50\,pc
resolution, CO is not detected. \cite{glover-2011} analyse
variations of the \aco factor as a function of visual extinction
in numerical simulations of low-metallicity clouds.
Following their study, our results are compatible with
mean visual extinctions of 0.1-1\,mag. This is within
values found from optical data \citep{izotov-1997,reines-2008}
and much lower than expected from IR data. The filling factor
of the shielded and IR-bright component in \sbs must be
very small. \cite{reines-2008} estimate a clumpy dust
covering factor of 60\,per cent.
With our 1-$\sigma$ limit of $W_{\rm CO(1-0)} \le 5$\,K\,\kms at
a resolution of 50\,pc, and a typical CO clump brightness of
15\,K\,\kms \citep{glover-2012}, we estimate a beam filling factor
for the CO-emitting clumps of less than 33\,per cent.

\section{Spectral energy distribution}
\label{sect:sed}

\begin{figure}
\centering
\includegraphics[clip,trim=5mm 1cm 0 0,width=8.2cm]{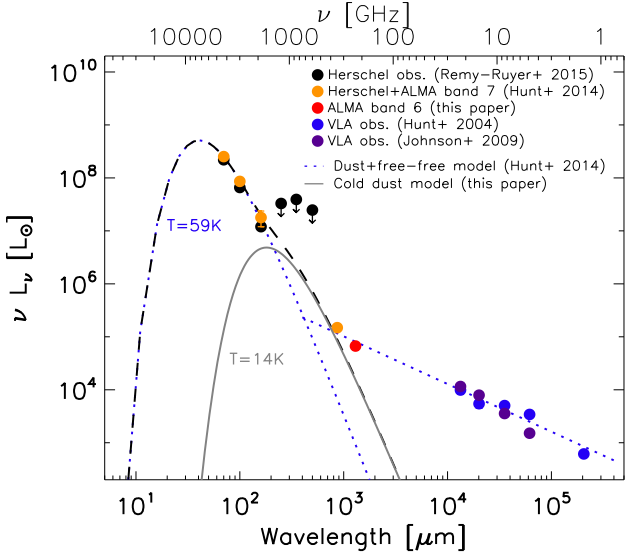}
\vspace{-15pt}
\caption{
Far-IR-radio Spectral Energy Distribution of \sbs.
The dashed black line shows the total dust model.
The individual model components and observations
are listed in the legend.
}
\label{fig:fig4}
\end{figure}

\begin{figure}
\centering
\includegraphics[clip,trim=5mm 1cm 0 0,width=7.5cm]{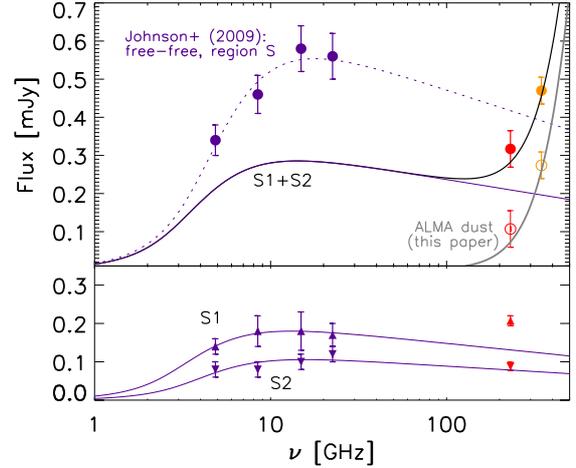}
\vspace{-5pt}
\caption{
\textit{Top:}
Radio SED of \sbs for region S with VLA observations from
\protect\cite{johnson-2009} in purple, and ALMA data from
\protect\cite{hunt-2014} in orange and from this paper in red.
The dotted purple line is the free-free model for region S
and the plain purple line is the summed free-free models
for S1 and S2, from \protect\cite{johnson-2009}.
The open circles are the ALMA data with free-free emission
from S1+S2 subtracted.
We model the ALMA data of region S with the free-free
models of S1+S2 and a dust component in grey. The full
model is shown in black.
\textit{Bottom:}
SED for the individual clusters S1 and S2 with VLA observations
and free-free models from \protect\cite{johnson-2009} in purple
and our ALMA data in red.
}
\label{fig:fig5}
\end{figure}

%
Figure~\ref{fig:fig4} shows the total FIR-radio SED of \sbs.
The photometry consists of \hers, ALMA, and VLA observations.
\cite{hunt-2014} model the ALMA 345\,GHz and VLA data with
free-free emission and the \hers data with warm dust
($T_{\rm dust}$=59\,K). Our ALMA observation (red point)
falls slightly below their free-free model.
Figure~\ref{fig:fig5} shows a zoom on the radio SED, focusing
on the high-resolution data. \cite{johnson-2009} model the
emission from S1, S2, and region S with three distinct free-free
components. For region S, their model passes in between
the ALMA band 6
and band 7 data but disagrees at a $>$2-$\sigma$ level from the
two observations. Moreover, the ALMA fluxes cannot be reproduced
simultaneously with free-free emission only. In particular,
the slope between the ALMA bands 7 and 6 is incompatible with
gas emission and is rather indicative of thermal dust emission.
For S2, the ALMA band 6 data match well the free-free
model of \cite{johnson-2009}.
For S1, our flux at 230\,GHz is 1.3 times higher than their
model (10-$\sigma$ significance) and we cannot fit both the
VLA and ALMA data within errors with a single free-free model.
Therefore in the following we consider an additional dust
component that is most visible in S1 and not S2.
We investigate the significance of a dust excess by focusing
on S1. Fitting the SED of S1 with an additional dust
component results in a better $\chi^2$ than without it,
and an F-test indicates that the probability of this
dust component is high ($>$98\,per cent).
This is valid when employing modified black-body spectra
as well as more realistic dust optical properties
and a distribution of temperatures \citep{galliano-2011}.

To model the dust emission present in the ALMA data
of region S, we subtract the free-free emission predicted
by the \cite{johnson-2009} models for the clusters S1 and S2.
Note that we do not subtract their free-free model for region S
(shown for reference in the top panel of Fig.~\ref{fig:fig5})
because there are discrepancies in the VLA observations/models
of \cite{johnson-2009} and \cite{hunt-2004} for region S and
the whole galaxy, respectively, whose origin is not clear to us.
Then, we fit a dust component to the ALMA excess of region S
(open circles in Fig.~\ref{fig:fig5}).
We find that the PACS data and ALMA excess
cannot be reproduced simultaneously
with a single modified black-body (MBB). This remains true
even if the PACS data encompass extended emission
not belonging to region S.
In addition to the MBB fitted to the PACS observations
(59\,K, $\beta=2$), we require a second dust component,
for which we assume common values of
$\kappa_{160}=1.6$\,m$^2$\,kg$^{-1}$ and
$\beta=1.7$ \citep{galliano-2011}.
The PACS 160\,\mum measurement puts an upper limit on the
dust temperature of $T_{\rm dust} \le 14$\,K, which gives a mass 
$M_{\rm dust}\simeq1.5\times10^5$\,\msun (grey lines in Figs.~\ref{fig:fig4}
and \ref{fig:fig5}). With a temperature of 10\,K, the required mass
is $\sim3\times10^5$\,\msun.
Those dust masses are about 4 times larger than the dust mass
of $3.8\times10^4$\,\msun found by \cite{hunt-2014} and
more than 10 times larger than previous dust estimates based
on \hers data only \citep{remy-2015}. We note that this cold
dust mass is an estimate only. In particular, alternative dust 
properties or mantle coating, resulting in a larger overall emissivity
would reduce the mass \citep[e.g.,][]{kohler-2014}. It is also possible
that the grains could exhibit an intrinsic excess at longer wavelength
\citep[e.g.,][]{meny-2007,coupeaud-2011}.
Further constraints between 200\,\mum and 600\,\mum would
be extremely valuable to confirm the cold dust hypothesis.

\section{Discussion}
\label{sect:disc}
\sbs is a very intriguing galaxy in the sense that it is actively
star-forming, metal-poor, with a large \hi reservoir and
extremely faint, still undetected CO emission. It potentially
harbours a large dust/hidden molecular reservoir.
Here we discuss the implications of such massive cold
reservoirs on the galaxy properties and theory of star-formation
(varying efficiencies, role of \hi in metal-poor environments).

Concerning the gas reservoir, with the SFR and $\tau_{\rm dep}$
values used in Sect.~\ref{sect:ks}, we find molecular
gas masses of $5\times10^6-8\times10^8$\,\msun.
This is the mass of gas within S1 and S2 required to explain
the current star formation.
If the depletion time is long ($\sim$2\,Gyr), the inferred molecular
mass is larger than the \hi mass within region S, estimated to be
around $4\times10^6-10^7$\,\msun by \cite{ekta-2009} and
\cite{hunt-2014}, and it is on the order of the total \hi mass of
$4\times10^8$\,\msun \citep{ekta-2009}.
If the depletion time is short ($\sim$20\,Myr), the inferred molecular
mass is of the order of the \hi mass in region S and lower than
the total \hi mass.
Recent numerical simulations at low metallicity demonstrate
that star formation can happen in a primarily atomic-dominated
ISM \citep{krumholz-2012b,glover-2016}. In the case of very short
depletion times, no hidden molecular gas is required in \sbs
to explain the SFR, while in the case of longer depletion times,
a massive amount of hidden molecular gas would be required.

Concerning the dust, the cold material found in Sect.~\ref{sect:sed}
could be responsible for the large extinction ($A_{V}\simeq20$\,mag)
needed in the mid-IR SED modeling. \cite{thuan-1999}
estimate the dust surface density to be 1.5\,M$_{\odot}$\,pc$^{-2}$.
Since the ALMA emission in S1 is barely resolved, we assume
that the cold dust spatial extent corresponds to our beam size.
Following the calculations of \cite{thuan-1999} and assuming a
uniform dust distribution, the cold dust mass needed to account
for the mid-IR extinction is $\sim5\times10^3$\,\msun, i.e. 30
times lower than our new estimate. This estimate will go up if
the dust is distributed in clumps, as suggested by \cite{reines-2008}.
It is noteworthy that the gas mass that is implied by our new
dust estimate, assuming full condensation of the metals into
dust, is of the same order of the gas mass estimated assuming
$\tau_{\rm dep}$ of 2\,Gyr.
This cold gas mass is a few times larger than the warm
\htwo mass of $\sim$10$^8$\,\msun (based on modeling
of the NIR \htwo lines; \citealt{thuan-2005a}), as found
in nearby metal-rich galaxies \citep[e.g.,][]{roussel-2007}.
Those facts further support our interpretation of an extreme
\aco value needed to explain the non-detection of CO emission
even with ALMA.

\section{Conclusion}
\label{sect:conc}
We present ALMA band\,6 (230\,GHz) observations in the
blue compact dwarf galaxy \sbs at an unprecedented resolution
of 0.2\,arcsec (52\,pc).
The CO\,$J$=2$\rightarrow$1 line is not detected. We report a
very deep 3-$\sigma$
$L_{\rm CO(2-1)}$ limit of $3.6\times10^4$\,\kkmspc.
This corresponds to a lower limit on the \aco
conversion factor of $10^2$\,\acounit when assuming
cloud depletion times and of $10^3 - 10^4$\,\acounit
for whole-galaxy depletion times.
The faintness of CO emission implies, for a whole-galaxy
star-formation depletion time, a dominant amount of CO-dark gas.
The ALMA continuum emission is detected and resolved over
the two main SSCs. Re-analysis of the IR-radio SED suggests
that the mm-fluxes are not only free-free emission but that
there is a cold dust component coincident with the position
of S1. Assuming standard dust properties, the dust mass could
be as large as $10^5$\,\msun.
This mass of cold dust corroborates the CO-dark gas interpretation.

\section*{Acknowledgements}
We acknowledge support from
the SYMPATICO grant (ANR-11-BS56-0023)
of the French Agence Nationale de la Recherche;
the DAAD/PROCOPE projects
57210883/35265PE;
the DFG for the programmes BI 1546/1-1 and HO 5475/2-1.
and for funding in SFB 881
'The Milky Way System' (subprojects B1, B2, and
B8) and in SPP 1573 'Physics of the ISM'
(grants KL 1358/18.1, KL 1358/19.2);
the ERC via the Advanced Grant 'STARLIGHT'
(project number 339177).
This paper makes use of the following ALMA data:
ADS/JAO.ALMA\#2013.1.00916.S. ALMA is a partnership
of ESO (representing its member states), NSF (USA) and
NINS (Japan), together with NRC (Canada), NSC and
ASIAA (Taiwan), and KASI (Republic of Korea), in cooperation
with the Republic of Chile. The Joint ALMA Observatory
is operated by ESO, AUI/NRAO and NAOJ.

\bibliographystyle{mnras}
\bibliography{mn-16-4176-arxiv}


\bsp	
\label{lastpage}
\end{document}